\documentclass[seceq]{ptptex}

\usepackage{graphicx}

\title{
Percolation Thresholds of the Fortuin-Kasteleyn Cluster
 for the Edwards-Anderson Ising Model
 on Complex Networks%
}

\subtitle{Analytical Results on the Nishimori Line}    

\author{
Chiaki \textsc{Yamaguchi}%
}

\inst{
Kosugichou 1-359, Kawasaki 211-0063, Japan
}

\recdate{
April 9, 2010; Revised July 14, 2010}

\abst{We analytically show the percolation thresholds
 of the Fortuin-Kasteleyn cluster for the Edwards-Anderson Ising model 
 on random graphs with arbitrary degree distributions.
 The results on the Nishimori line are shown.
 We obtain the results for the $\pm J$ model,
 the diluted $\pm J$ model, 
 and the Gaussian model,
 by applying an extension of a criterion
 for the random graphs
 with arbitrary degree distributions.
 The results for the infinite-range $\pm J$ model 
 and the Sherrington-Kirkpatrick model are also shown.}

\begin{document}
\maketitle

\section{Introduction}

The study of complex networks has been carried out, and
 the study of spin models on the complex networks
 is important \cite{DGM1}.
 As an example of such a spin model, 
 we study in this article spin models on
 random graphs with arbitrary degree distributions. 
 The behavior of spins on a no growing network is investigated.

We investigate the Edwards-Anderson Ising model \cite{EA}
 as an Ising spin-glass model.
 The understanding of the Edwards-Anderson Ising model
 on random graphs and on the Bethe lattice is still incomplete \cite{DGM1, VB, MP}.
 In this article, 
 the $\pm J$ model, the diluted $\pm J$ model,
 and the Gaussian model for the Edwards-Anderson Ising model
 are investigated.
 For those models,
 there is a special line, called the Nishimori line,
 on the phase diagram 
 for the exchange interactions and the temperature.
 The internal energy, the upper bound of the specific heat, 
 and so forth are exactly calculated on the Nishimori line
  \cite{N1, MH, N2, HM, H}.
 The location of the multicritical point for the Edwards-Anderson Ising model
 on the square lattice is conjectured,
 and it is shown that 
 the conjectured value is in good agreement
 with other numerical estimates 
 \cite{NN}.
 In this article, the results on the Nishimori line
  are shown.

There is a case where
 a percolation transition of networks occures.
 A network is
 divided into many networks
 by deleting some of its nodes and/or links.
 We call this percolation transition `the percolation transition
 of network' in this article.
 There is also a case where a percolation transition
 of clusters occurs. A cluster consists of fictitious bonds.
 The bond is put between spins.
 One of the clusters becomes a giant component when a cluster is percolated.
 We call this percolation transition `the percolation transition
 of clusters', and discuss
 the percolation transition of a cluster
 on a complex network.

In this article,
 the percolation transition of the Fortuin-Kasteleyn (FK)
 cluster is investigated.
 The FK cluster has the FK representation \cite{KF, FK}.
 In the ferromagnetic Ising model,
 the percolation transition point
 agrees with the phase transition point.
 The agreement is described
 in Ref.~\citen{CK}. 
 Powerful Monte Carlo methods using the FK cluster
 have been proposed \cite{SW, KG, TO, YK, YKO}.
 The Edwards-Anderson Ising model has a conflict
 in the interactions:
 the percolation transition point 
 disagrees with the phase transition point.
 There are numerous approches for resolving the disagreement
 by extending the FK representation \cite{MNS}.
 On the other hand, it was pointed out
 by de Arcangelis et al.
 that the correct understanding of the percolation phenomenon
 of the FK cluster in the Edwards-Anderson Ising model
 is important since a dynamical transition 
 occurs at a temperature very close to the percolation
 temperature, and the dynamical transition and
 percolation transition are related to
 a transition for a signal propagating between spins \cite{ACP}.
 The dynamical transition is characterized by a parameter
 called the Hamming distance or damage \cite{ACP}.
 In this article,
 the percolation threshold is analytically found.

The study of random graphs with arbitrary degree distributions
 has been carried out~\cite{Ne}.
 Our results are obtained
 by applying an extension of a criterion \cite{MR, CEAH, NSW}
 for random graphs with arbitrary
 degree distributions.
 The results for the infinite-range $\pm J$ model
 and the Sherrington-Kirkpatrick (SK) model \cite{SK} 
 are also shown.

This article is organized as follows.
 First in \S\ref{sec:2},
 a complex network model and the Edwards-Anderson Ising model
 are described. 
 Next in \S\ref{sec:3},
 the FK cluster is explained.
 After elucidating
 a criterion for the percolation of a cluster
 in \S\ref{sec:4},
 we will in \S\ref{sec:5} find the percolation thresholds 
 for the $\pm J$ model and the diluted $\pm J$ model.
 The result for the Gaussian model is shown in \S\ref{sec:6}.
 The final section is devoted to a summary.

\section{A complex network model and the Edwards-Anderson Ising model} \label{sec:2}

A network consists of nodes and links.
 A link connects nodes.
 In this article, as a complex network model, 
 random graphs with arbitrary degree distributions are investigated.
 The network has no correlation between nodes.
 The node degree, $k$, is given with a distribution $p (k)$. 
 The links are randomly connected between nodes.

We define a variable $b (i, j)$, where
 $b (i, j)$ is one when nodes $i$ and $j$
 are connected by a link.
 $b (i, j)$ is zero when
 nodes $i$ and $j$ are not connected by a link.
 The degree $k (i)$ of node $i$ is given by
\begin{equation}
 k (i) = \sum_j b (i, j) \, .
\end{equation}
 The coordination number (the average of the node degree for links), 
 $\langle k \rangle _N$, is given by 
\begin{equation}
 \langle k \rangle_N = \frac{1}{N} \sum_i^N k (i) \, ,
\end{equation}
 where $\langle \, \rangle_N$ is the average over the entire network.
 $N$ is the number of nodes.
 The average of the square of the node degree for links, 
 $\langle k^2 \rangle_N$, is given by
\begin{equation}
 \langle k^2 \rangle_N = \frac{1}{N} \sum_i^N k^2 (i) \, .
\end{equation}

We define 
\begin{equation}
 a = \frac{2 \langle k \rangle_N}{\langle k^2 \rangle_N} \, , \label{eq:a}
\end{equation}
 where $a$ represents an aspect of the network.

\begin{figure}[t]
\begin{center}
\includegraphics[width=0.95\linewidth]{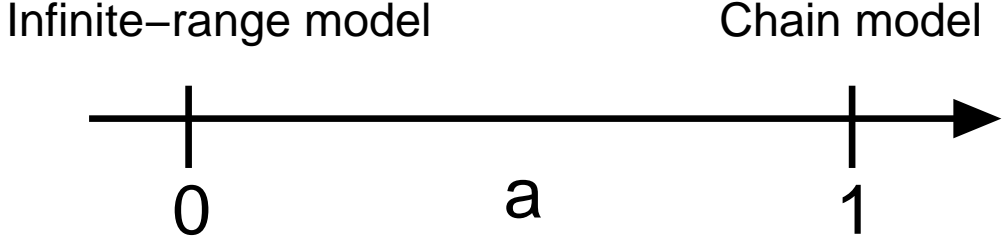}
\end{center}
\caption{
 Relation between the aspect $a$ and the model on the network.
\label{fig:a}
}
\end{figure}

Figure \ref{fig:a} shows the
 relation between the aspect $a$ and the model on the network.
 The network is almost a complete graph when $a$
 is close to zero, and
 the model on the network is almost an infinite-range model.
 The model on the network is the infinite-range model 
 when $\langle k \rangle_N = N - 1$, $\langle k^2 \rangle_N = (N - 1)^2$, and $a = 2 / (N - 1)$.
 The network consists of many cycle graphs
 when the coordination number $\langle k \rangle_N$ is two.
 The model on the network consists of many chain models when $\langle k \rangle_N$ is two.
 In the Erd\H{o}s-R\'enyi (ER) random graph model and in the Gilbert model,
 the distribution of node degree is the Poisson distribution \cite{DGM1}.
 The ER random graph model is a network model wherein the network
 consists of a fixed number of nodes and
 a fixed number of links, and the links are randomly connected between the nodes.
 The Gilbert model is a network model wherein the
 link between nodes is connected with a given probability.
 In the ER random graph model and in the Gilbert model,
 $\langle k \rangle_N = 1$ and $\langle k^2 \rangle_N = \langle k \rangle_N (\langle k \rangle_N + 1) = 2$ when 
 $a$ is one.

The Hamiltonian for the Edwards-Anderson Ising model, ${\cal H}$, 
 is given by
\begin{equation}
 {\cal H} = -  \sum_{\langle i, j \rangle} J_{i, j} S_i S_j \, ,
\end{equation}
 where $\langle i, j \rangle$ denotes nearest-neighbor pairs, $S_i$ denotes
 the state of the spin at node $i$, and $S_i = \pm 1$.
 $J_{i, j}$ is the strength of the exchange interaction between spins.
 The value of $J_{i, j}$ is given by the distribution $P (J_{i, j})$.
 The $\pm J$ model, the diluted $\pm J$ model, and the Gaussian model
 are given by specific $P (J_{i, j})$.

For the $\pm J$ model, the distribution $P^{(\pm J )} (J_{ij})$
 is given by
\begin{equation}
 P^{(\pm J)} (J_{i, j})
 = p \, \delta_{J_{i, j}, J} + (1 - p) \, \delta_{J_{i, j}, - J} \, , \label{eq:PpmJJij}
\end{equation}
 where $J > 0$. 
 $p$ is the probability that the interaction
  is ferromagnetic ($J_{i, j} = J$).
 $1 - p$ is the probability that the interaction
 is antiferromagnetic ($J_{i, j} = - J$).

For the diluted $\pm J$ model, the distribution $P^{({\rm D} \pm J)} (J_{i, j})$
 is given by
\begin{equation}
 P^{({\rm D} \pm J)} (J_{i, j})
 = p \, \delta_{J_{i, j}, J} + q \, \delta_{J_{i, j}, - J}
 + r \, \delta_{J_{i, j}, 0} \, , \label{eq:PDpmJJij}
\end{equation}
 where $J > 0$ and $p + q + r = 1$.
 $p$ is the probability that the interaction
 is ferromagnetic ($J_{i, j} = J$).
 $q$ is the probability that the interaction
 is antiferromagnetic ($J_{i, j} = - J$).
 $r$ is the probability that the interaction
 is diluted ($J_{i, j} = 0$).
 This model is the $\pm J$ model when $r = 0$.

For the Gaussian model, the distribution
 $P^{({\rm Gaussian})} (J_{i, j})$ is given by
\begin{equation}
 P^{(\rm Gaussian)} (J_{i, j}) = \frac{1}{\sqrt{2 \pi J^2}}
 e^{- (J_{i, j} - J_0 )^2 / 2 J^2} \, . \label{eq:PGaussJij}
\end{equation}
 The average of $J_{i, j}$ is given by $[ J_{i, j} ]_R = J_0$, where
 $[ \, ]_R$ is the random configuration average.
 The variance of $J_{i, j}$ is given by $[ J^2_{i, j} ]_R - [ J_{i, j} ]^2_R = J^2$.

To calculate thermodynamic quantities,
 a gauge transformation \cite{N1, MH, N2, HM, H, T}
 wherein the transformation is 
 performed by
\begin{equation}
 J_{i, j} \to J_{i, j} \sigma_i \sigma_j \, , \quad S_i \to S_i \sigma_i 
 \label{eq:GaugeT} 
\end{equation}
 is used, where $\sigma_i = \pm 1$.
 It is known that the gauge transformation has no effect
 on thermodynamic quantities \cite{T}.
 Following the gauge transformation,
 the ${\cal H}$ part becomes ${\cal H} \to {\cal H}$ and
 the $P (J_{i, j} )$ part becomes $P (J_{i, j} ) \to P (J_{i, j} \sigma_i \sigma_j)$.

\section{The Fortuin-Kasteleyn cluster} \label{sec:3}

The bond for the FK cluster is put between spins
 with probability $P_{\rm FK} (S_i, S_j, J_{ij})$.
 The value of $P_{\rm FK}$ depends on
 the interaction between spins and the states of spins.
 We call the bond the FK bond in this article.
 $P_{\rm FK} (S_i, S_j, J_{ij})$ is given by \cite{ACP}
\begin{equation}
 P_{\rm FK} (S_i, S_j, J_{ij}) = 1 -
 e^{- \beta J_{ij} S_i S_j - \beta |J_{ij}|} \, , \label{eq:PbondSiSjJij}
\end{equation}
 where $\beta$ is the inverse temperature and $\beta = 1 / k_B T$.
 $k_B$ is the Boltzmann constant and $T$ is the temperature.
 By connecting the FK bonds, the FK clusters are generated.
 By the gauge transformation,
 the $P_{\rm FK}$ part becomes $P_{\rm FK} \to P_{\rm FK}$.

\begin{figure}[t]
\begin{center}
\includegraphics[width=0.65\linewidth]{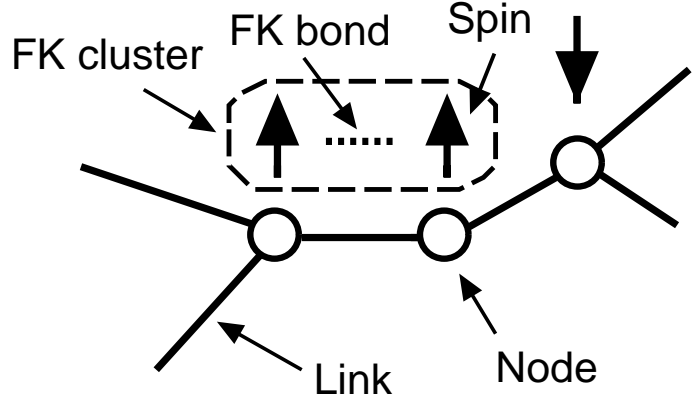}
\end{center}
\caption{
 Network and FK cluster.
 Three nodes, six links, three spins, an FK bond, and an FK cluster are depicted.
 Spins are aligned on each node.
 The percolation of the FK cluster is discussed in this article.
 \label{fig:network-cluster}
}
\end{figure}
 Figure \ref{fig:network-cluster} shows a conceptual diagram
 of a network and an FK cluster.
 Three nodes, six links, three spins, an FK bond, and an FK cluster are depicted.
 Spins are aligned on each node.
 The percolation of the FK cluster is discussed in this article.

The thermodynamic quantity of the FK bond put between
 the spins on nodes $i$ and $j$,
 $[\langle b_{\rm FK} (i, j) \rangle_T ]_R$, is given by
\begin{equation}
 [\langle b_{\rm FK} (i, j) \rangle_T ]_R =
 [\langle P_{\rm FK} (S_i, S_j, J_{i, j}) \rangle_T ]_R \, , \label{eq:bbondijTR}
\end{equation}
 where $\langle \, \rangle_T$ is the thermal average.
  The thermodynamic quantity of the node degree for FK bonds at node $i$,
 $[\langle k_{\rm FK} (i) \rangle_T ]_R$, is given by
\begin{eqnarray}
 & & [\langle k_{\rm FK} (i) \rangle_T ]_R = \nonumber \\
 & &  \quad [\langle \sum_{\{ j | b(i, j) = 1\}}
 P_{\rm FK} (S_i, S_j, J_{i, j}) \rangle_T ]_R \, . \label{eq:kbondiTR}
\end{eqnarray}
 The thermodynamic quantity of the square of the node degree for FK bonds at node $i$,
 $[\langle k^2_{\rm FK} (i) \rangle_T ]_R$, is given by
\begin{eqnarray}
 & & [\langle k^2_{\rm FK} (i) \rangle_T ]_R = \nonumber \\
 & & [\langle \sum_{\{ j | b(i, j) = 1\}}
 \sum_{\{ l | b(i, l) = 1\}} P_{\rm FK} (S_i, S_j, J_{i, j})
  \nonumber \\ 
 &\times& P_{\rm FK} (S_i, S_l, J_{i, l}) (1 - \delta_{j, l})   
  \nonumber \\
 &+& \sum_{\{ j | b(i, j) = 1\}} P_{\rm FK} (S_i, S_j, J_{i, j}) \rangle_T ]_R \, . 
 \label{eq:k2bondiTR}
\end{eqnarray}
 The thermodynamic quantity of the node degree for FK bonds, $[\langle k_{\rm FK} \rangle_T ]_R$,
 is given by
\begin{equation}
 [\langle k_{\rm FK} \rangle_T ]_R = \frac{1}{N} \sum_i^N
 [\langle k_{\rm FK} (i) \rangle_T ]_R \, . \label{eq:kbondTR} 
\end{equation}
 The thermodynamic quantity of the square of the node degree for FK bonds,
 $[\langle k^2_{\rm FK} \rangle_T ]_R$, is given by
\begin{equation}
 [\langle k^2_{\rm FK} \rangle_T ]_R = \frac{1}{N} \sum_i^N
 [\langle k^2_{\rm FK} (i) \rangle_T ]_R \, . \label{eq:k2bondTR}
\end{equation}

\section{A criterion for percolation of clusters} \label{sec:4}

The percolation of the random graphs with
 arbitrary degree distributions occurs when \cite{MR, CEAH, NSW}
\begin{equation}
 \langle k^2 \rangle_N \, \ge \, 2 \langle k \rangle_N  \, . \label{eq:k2k1}
\end{equation}
 Equation~(\ref{eq:k2k1}) is the inequality when 
 the network is percolated.
 Equation~(\ref{eq:k2k1}) is the equality when 
 the network is at the percolation transition point.
 The criterion (Eq.~(\ref{eq:k2k1})) is true
 for a sufficiently large number of nodes.
 Equation~(\ref{eq:k2k1}) is derived
  by Molly and Reed \cite{MR},
 Cohen et al. \cite{CEAH},
 and Newman et al. \cite{NSW}.

From Eq.~(\ref{eq:k2k1}),
 the network is percolated when $a < 1$ and
 the network is at the percolation transition point when $a = 1$.
 The network is unpercolated when $a > 1$.
 Therefore, the percolation of clusters
 is investigated for $0 < a \le 1$.

When links and/or nodes are randomly diluted
 on the random graphs with arbitrary degree distributions,
 the criterion (Eq.~(\ref{eq:k2k1}))
 is applicable to the diluted network \cite{CEAH}.
 The percolation problem for the diluted network can be regarded
 as the random-bond percolation problem.
 We define the bond states as the graph $G$.
 We define the node degree for random bonds at node $i$ as $k_{\rm random \, bond} (G, i)$.
 The random bonds are randomly put on the links, and the links are randomly connected
 between the nodes.
 The criterion of the percolation of clusters for the random-bond percolation problem
 is given by
\begin{equation}
  \frac{1}{N} \sum_{i} k_{\rm random \, bond}^2 (G, i) \ge \frac{2}{N}  \sum_i 
 k_{\rm random \, bond} (G, i) \, . \label{eq:k2k1G_cohen}
\end{equation}

In what follows, 
 a criterion of the percolation of clusters for spin models
 is conjectured on the basis of the above discussion.
 
We consider a case that
 the magnitude of a bond does not depend on the degree $k (i)$.
 The bond is a bond put between spins
 and includes the FK bond.
 We define a variable 
 for the inverse temperature as $\rho (\beta )$.
 We set
\begin{equation}
 0 < \rho (\beta ) \le 1 \, . \label{eq:rhorange}
\end{equation}
 We consider a case that 
 $[\langle b_{\rm bond} (i, j) \rangle_T]_R$,
 $[\langle k_{\rm bond } (i) \rangle _T]_R$, and 
 $[\langle k^2_{\rm bond} (i) \rangle_T ]_R$ are respectively written as
\begin{equation}
 [\langle b_{\rm bond} (i, j) \rangle_T ]_R =
 \rho (\beta )  \, , \label{eq:bondij}
\end{equation}
\begin{equation}
 [\langle k_{\rm bond} (i) \rangle_T ]_R =
 \rho (\beta ) \, k (i) \, , \label{eq:k1bondi}
\end{equation}
\begin{eqnarray}
 [\langle k^2_{\rm bond} (i) \rangle_T ]_R &=& 
 \rho^2 ( \beta ) \, k (i) 
 [ k (i) - 1 ]  \nonumber \\
 &+& \rho ( \beta ) \, k (i) \, .
 \label{eq:k2bondi}
\end{eqnarray}
 In this case, it is implied that
 the bias for $k (i)$ does not appear
 in the statistical results of the bonds.
 Therefore, we describe the case that
 $[\langle b_{\rm bond} (i, j) \rangle_T]_R$,
 $[\langle k_{\rm bond } (i) \rangle _T]_R$, and 
 $[\langle k^2_{\rm bond} (i) \rangle_T ]_R$ are respectively written as 
 Eqs.~(\ref{eq:bondij}), (\ref{eq:k1bondi}), and (\ref{eq:k2bondi})
 as the case that the magnitude of the bond does not depend on $k (i)$.

When the magnitude of the bond does not depend on $k (i)$,
 as an extension of Eq.~(\ref{eq:k2k1G_cohen}),
 we conjecture
\begin{eqnarray}
 & & \frac{1}{N} \sum_{i} k_{\rm bond}^2 (\{ S_j \},
 \{J_{j, l} \}, G, i) \ge \nonumber \\
 & &  \quad \frac{2}{N}  \sum_i 
 k_{\rm bond} (\{ S_j \},\{ J_{j, l} \}, G, i) \, . \label{eq:k2k1G}
\end{eqnarray}
 Since the magnitude of the bond does 
 not depend on $k (i)$, the bonds are randomly put 
 on links, and
 the links are randomly connected between nodes.
 $\{ S_j \}$ is a subset of spin states.
 $\{ J_{j, l} \}$ is a subset of exchange interactions.
 $k_{\rm bond} (\{ S_j \},
 \{ J_{j, l} \}, G, i)$ is the node degree for bonds
 at node $i$ in the graph $G$
 that is compatible with $\{ S_j \}$ and $\{ J_{j, l} \}$.
 Equation~(\ref{eq:k2k1G_cohen}) is true for a sufficiently large number of nodes.
 Therefore, Eq.~(\ref{eq:k2k1G}) may also be 
 true for a sufficiently large number of nodes when
 the magnitude of the bond does not depend on $k (i)$.
 By using Eq.~(\ref{eq:k2k1G}),
 we obtain a conjectured criterion of the percolation of clusters for spin models
 as
\begin{equation}
 [\langle k^2_{\rm bond} \rangle_T ]_R
 \ge 2 [\langle k_{\rm bond} \rangle_T ]_R \, .
 \label{eq:k2k1bond}
\end{equation}
Equation~(\ref{eq:k2k1bond}) is the inequality when 
 the cluster is percolated.
 Equation~(\ref{eq:k2k1bond}) is the equality when 
 the cluster is at the percolation transition point. 
 Equation~(\ref{eq:k2k1bond}) 
 gives the percolation threshold of the clusters.

\section{The $\pm J$ model and the diluted $\pm J$ model} \label{sec:5}

 By using Eq.~(\ref{eq:PpmJJij}), the distribution $P^{(\pm J)} (J_{i, j})$ 
 is written as 
\begin{equation}
 P^{(\pm J)} (J_{i, j}) = \frac{e^{\beta_P J_{i, j}}}
{2 \cosh (\beta_P J)} \, , 
 \quad J_{i, j} = \pm J \, , \label{eq:PpmJJij2}
\end{equation}
 where $\beta_P$ is given by \cite{N1, MH, N2, HM}
\begin{equation}
 \beta_P = \frac{1}{2 J} \ln \frac{p}{1-p} \, . \label{eq:betaPpmJ}
\end{equation}
 When the value of $\beta_P$
 is consistent with the value of 
 the inverse temperature $\beta$, 
 the line on the phase diagram obtained using Eq.~(\ref{eq:betaPpmJ})
 is called the Nishimori line.
 By using the gauge transformation,  
 the distribution $P^{(\pm J)} (J_{i, j})$ part
 becomes
\begin{eqnarray}
 \prod_{\langle i, j \rangle} P^{(\pm J)} (J_{i, j})
 &=& \frac{e^{\beta_P \sum_{\langle i, j \rangle} J_{i, j}}}
{[2 \cosh (\beta_P J)]^{N_B}}  \nonumber \\
 &\to& \frac{\sum_{\{ \sigma_i \}} e^{\beta_P \sum_{\langle i, j \rangle}
 J_{i, j} \sigma_i \sigma_j}}
 {2^ N [2 \cosh (\beta_P J)]^{N_B}} \, ,
 \label{eq:PpmJJij3}
\end{eqnarray}
 where $N_B$ is the number of nearest-neighbor pairs
 in the whole system.

 By using Eqs.~(\ref{eq:GaugeT}), (\ref{eq:PbondSiSjJij}), (\ref{eq:bbondijTR}),
 and (\ref{eq:PpmJJij3}), when $\beta = \beta_P$, 
 the thermodynamic quantity of the FK bond put between
 the spins on nodes $i$ and $j$,
 $[\langle b_{\rm FK} (i, j) \rangle_T ]^{(\pm J)}_R$, is obtained as
\begin{eqnarray}
 & & [\langle b_{\rm FK} (i, j ) \rangle_T ]^{(\pm J)}_R \nonumber \\
 &=& \sum_{ \{ J_{l, m} \}}
 \prod_{\langle l, m \rangle} P^{(\pm J )} (J_{l, m})  \nonumber \\
 &\times& 
  \frac{\sum_{\{ S_l \} } 
 P_{\rm FK} (S_i, S_j, J_{i, j})
 \, e^{\beta_P \sum_{\langle l, m \rangle} J_{l, m} S_l S_m}}
 {\sum_{\{ S_l \} } 
 e^{\beta_P \sum_{\langle l, m \rangle} J_{l, m} S_l S_m}} \nonumber \\
  &=& \frac{1}{2^N [2 \cosh (\beta_P J)]^{N_B}}  \nonumber \\
 &\times& \sum_{ \{ J_{l, m} \}}
 \sum_{\{ S_l \} }
 P_{\rm FK} (S_i, S_j, J_{i, j})
 \, e^{\beta_P \sum_{\langle l, m \rangle} J_{l, m} S_l S_m} \nonumber \\ 
 &=& \tanh ( \beta_P J) \, .  
 \label{eq:bbondijpmJ}
\end{eqnarray}
 By using
 Eqs.~(\ref{eq:GaugeT}), (\ref{eq:PbondSiSjJij}), (\ref{eq:kbondiTR}),
 and (\ref{eq:PpmJJij3}), when $\beta = \beta_P$, 
 the thermodynamic quantity of the node degree for FK bonds at node $i$,
 $[\langle k_{\rm FK} (i) \rangle_T ]^{(\pm J)}_R$, is obtained as 
\begin{equation}
 [\langle k_{\rm FK} (i) \rangle_T ]^{(\pm J)}_R =
 \tanh ( \beta_P J) 
 \, k (i) \, . \label{eq:kbondipmJ}
\end{equation}
 By using
 Eqs.~(\ref{eq:GaugeT}), (\ref{eq:PbondSiSjJij}), (\ref{eq:k2bondiTR}),
 and (\ref{eq:PpmJJij3}),
 when $\beta = \beta_P$, 
 the thermodynamic quantity of the square of the node degree for FK bonds at node $i$,
 $[\langle k^2_{\rm FK} (i) \rangle_T ]^{(\pm J)}_R$, is obtained as 
\begin{eqnarray}
 [\langle k^2_{\rm FK} (i) \rangle_T ]^{(\pm J)}_R &=& 
  \tanh^2 ( \beta_P J) \,  k (i) 
 [ k (i) - 1 ] 
 \nonumber \\
 &+& 
  \tanh ( \beta_P J) \, k (i) \, .
 \label{eq:k2bondipmJ}
\end{eqnarray}

We set
\begin{equation}
 \rho^{(\pm J)}  ( \beta_P ) = 
  \tanh ( \beta_P J) \, . \label{eq:rhopmj}
\end{equation}
 Equations~(\ref{eq:bbondijpmJ}),
 (\ref{eq:kbondipmJ}), (\ref{eq:k2bondipmJ}), and (\ref{eq:rhopmj}) are
 formulated as Eqs.~(\ref{eq:bondij}), (\ref{eq:k1bondi}), 
 and (\ref{eq:k2bondi}).
 Therefore, the magnitude of the FK bond does not depend on $k (i)$.
 By using Eqs.~(\ref{eq:kbondTR}), (\ref{eq:k2bondTR}), (\ref{eq:k2k1bond}),
 (\ref{eq:kbondipmJ}), and (\ref{eq:k2bondipmJ}), we obtain
\begin{equation}
 1- \exp (- 2 \beta_P J)
 \ge \frac{2 \langle k \rangle_N}{\langle k^2 \rangle_N} \, . \label{eq:bondPcpmJ}
\end{equation}
 Equation~(\ref{eq:bondPcpmJ}) is the inequality when 
 the FK cluster is percolated.
 Equation~(\ref{eq:bondPcpmJ}) is the equality when 
 the FK cluster is at the percolation transition point. 

From Eqs.~(\ref{eq:rhorange}) and (\ref{eq:rhopmj}),
 there is the percolation transition point for $0 < \beta_P \le \infty$.
 From Eq.~(\ref{eq:bondPcpmJ}),
 there is the percolation transition point for $0 < a \le 1$.
 By using Eqs.~(\ref{eq:betaPpmJ}) and (\ref{eq:bondPcpmJ}),
 the probability $p^{(\pm J)}$ that 
 the interaction is ferromagnetic is obtained as
\begin{equation}
 p^{(\pm J)} = \frac{1}{2 - a}  \label{eq:pbondPmJ}
\end{equation}
 at the percolation transition point.
 By using Eqs.~(\ref{eq:betaPpmJ}) and (\ref{eq:pbondPmJ}), 
 the percolation transition temperature $T^{(\pm J)}_P$ is obtained as
\begin{equation}
 T^{(\pm J )}_P = \frac{J}{k_B} \frac{2}{\ln \frac{1}{1 - a}} \, . \label{eq:TPbondPmJ}
\end{equation}

\begin{figure}[t]
\begin{center}
\includegraphics[width=0.95\linewidth]{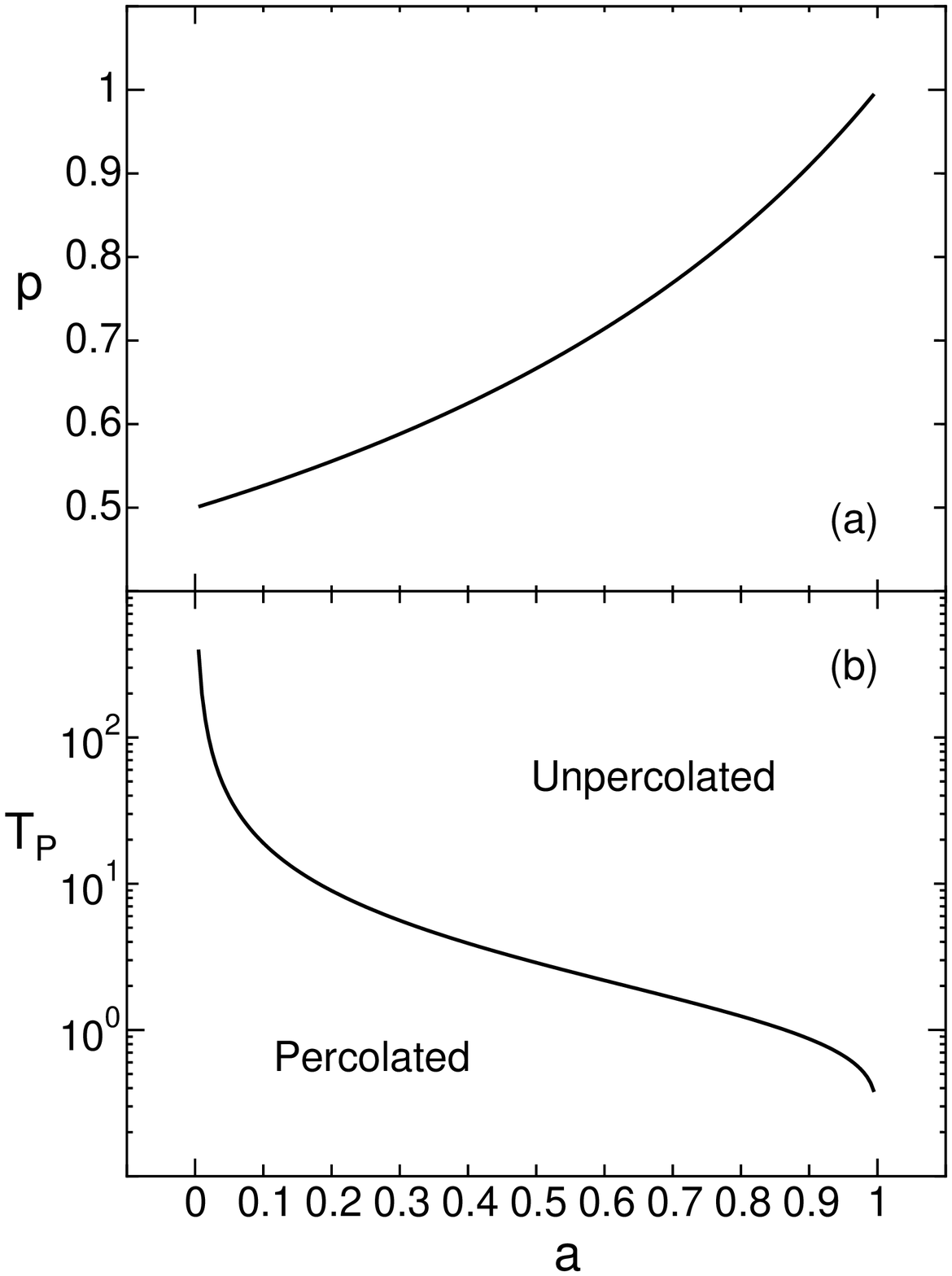}
\end{center}
\caption{
 Percolation threshold of the FK cluster for the $\pm J$ model.
 (a) The relation between the aspect $a$ and the probability
 $p^{(\pm J)}$ is shown.
 (b) The relation between the aspect $a$ and the percolation
 transition temperature $T^{(\pm J)}_P$ is shown.
 $J / k_B$ is set to $1$.
\label{fig:pmJ}
}
\end{figure}
Figure~\ref{fig:pmJ} shows
 the percolation threshold of the FK cluster for the $\pm J$ model.
 Figure~\ref{fig:pmJ}(a) shows
 the relation between the aspect $a$ and the probability
 $p^{(\pm J)}$.
 Equation~(\ref{eq:pbondPmJ}) is used for showing Fig.~\ref{fig:pmJ}(a).
 Figure~\ref{fig:pmJ}(b) shows
 the relation between the aspect $a$ and the percolation
 transition temperature $T^{(\pm J)}_P$.
 Equation~(\ref{eq:TPbondPmJ}) is used for showing Fig.~\ref{fig:pmJ}(b).
 $J / k_B$ is set to $1$.

For the ferromagnetic Ising model on the same network,
 the phase transition temperature $T^{({\rm Ferro})}_C$ is
 \cite{LVVZ, DGM2}
\begin{equation}
 T^{({\rm Ferro})}_C
 = \frac{J}{k_B} \frac{2}{\ln \frac{1}{1 - a}} \, . 
\end{equation}
 $T^{(\pm J)}_P$ (Eq.~(\ref{eq:TPbondPmJ}))
 coincides with 
 $T^{({\rm Ferro})}_C$.

The complete graph is considered as $a \sim 0$.
 We set
 $\langle k \rangle_N = N - 1$, $\langle k^2 \rangle_N = (N - 1)^2$,
 $a = 2 / (N - 1)$, and $J \to J / \sqrt{N}$.
 From the settings, the model on the network becomes the infinite-range $\pm J$ model.
 By using Eq.~(\ref{eq:pbondPmJ}),
 the probability $p^{({\rm IR} \pm J)}$ that 
 the interaction is ferromagnetic is obtained as 
\begin{equation}
 p^{({\rm IR} \pm J)} = \frac{N - 1}{2 (N - 2)} \to \frac{1}{2}  \label{eq:pbondPmJ2}
\end{equation}
 for a sufficiently large number of nodes
 at the percolation transition point.
 By using Eq.~(\ref{eq:TPbondPmJ}),
 the percolation transition temperature $T^{({\rm IR} \pm J)}_P$
 is obtained as
\begin{equation}
 T^{({\rm IR} \pm J )}_P =
 \frac{J}{k_B \sqrt{N}} \frac{2}{\ln ( 1 + \frac{2}{N - 3})} 
 \to \frac{J}{k_B} \sqrt{N}  \label{eq:TPbondPmJ2}
\end{equation}
 for a sufficiently large number of nodes.

In Ref.~\citen{MNS},
 the percolation transition temperature of the FK cluster 
 for the infinite-range $\pm J$ model is derived
 by using the analytical solution of the SK model.
 The percolation transition temperature of the FK cluster
 for the infinite-range $\pm J$ model obtained in this article
 agrees with the result for a single-replica case in Ref.~\citen{MNS}.
 Therefore, we were able to confirm 
 that our result is exact at this extremal point. 

We consider the case for $a = 1$.
 By using Eq.~(\ref{eq:pbondPmJ}), we obtain $p = 1$.
 By using Eq.~(\ref{eq:TPbondPmJ}), we obtain $T_P = 0$.
 From Eq.~(\ref{eq:k2k1}),
 the network is at the percolation transition point.
 From $p = 1$, the exchange interaction is only the ferromagnetic interaction.
 From $T_P = 0$, all the spins are parallel.
 From $p = 1$ and $T_P = 0$, we obtain $P_{\rm FK} = 1$
 for all nearest-neighbor pairs.
 Therefore, the FK cluster and the network are 
 at the percolation transition point.
 We were able to confirm 
 that our result is exact at this extremal point. \\

 By using Eq.~(\ref{eq:PDpmJJij}), 
 the distribution $P^{({\rm D} \pm J )} (J_{ij})$ for
 the diluted $\pm J$ model is written as
\begin{equation}
 P^{({\rm D} \pm J)} (J_{i, j}) =
 \frac{e^{\beta^{(2)}_P J^2_{i, j} + \beta_P J_{ij}}}
 {e^{\beta^{(2)}_P J^2 + \beta_P J} + 1 + e^{\beta^{(2)}_P J^2 - \beta_P J}} 
 \, , \label{eq:PJijdpmJ}
\end{equation}
 where $\beta_P^{(2)}$ and $\beta_P$ are respectively \cite{H}
\begin{equation}
 \beta^{(2)}_P = \frac{1}{J^2} \ln \sqrt{\frac{pq}{r^2}}
 \, , \quad \beta_P = \frac{1}{J} \ln \sqrt{\frac{p}{q}} \, . \label{eq:betadpmJ}
\end{equation}
 This model becomes the $\pm J$ model when $r = 0$.
 In what follows, the result for $r \ne 0$ is only described
 since the result for the $\pm J$ model is described above.
 By using the gauge transformation,  
 the distribution $P^{({\rm D} \pm J)} (J_{i, j})$ part
 becomes
\begin{eqnarray}
 & & \prod_{\langle i, j \rangle} P^{({\rm D} \pm J)} (J_{i, j}) \nonumber \\
 &=& 
 \frac{e^{\beta^{(2)}_P \sum_{\langle i, j \rangle} J^2_{i, j} + \beta_P \sum_{\langle i, j \rangle} J_{i, j}}}
 {(e^{\beta^{(2)}_P J^2 + \beta_P J} + 1 + e^{\beta^{(2)}_P J^2 - \beta_P J})^{N_B}} 
 \nonumber \\  &\to& 
 \frac{\sum_{\{ \sigma_i \}} 
 e^{\beta^{(2)}_P \sum_{\langle i, j \rangle} J^2_{i, j} + \beta_P \sum_{\langle i, j \rangle} J_{i, j}
  \sigma_i \sigma_j}}
 {2^N (e^{\beta^{(2)}_P J^2 + \beta_P J} + 1 + e^{\beta^{(2)}_P J^2 - \beta_P J})^{N_B}} 
 \, .
 \label{eq:PJijdpmJ2}
\end{eqnarray}

By using
 Eqs.~(\ref{eq:GaugeT}), (\ref{eq:PbondSiSjJij}), (\ref{eq:bbondijTR}),
 and (\ref{eq:PJijdpmJ2}), when $\beta = \beta_P$, 
 the thermodynamic quantity of the FK bond put between
 the spins on nodes $i$ and $j$,
 $[\langle b_{\rm FK} (i, j) \rangle_T ]^{({\rm D} \pm J)}_R$,
 is obtained as
\begin{eqnarray}
 & & [ \langle b_{\rm FK} (i, j) \rangle_T ]^{({\rm D} \pm J)}_R \nonumber \\
 &=& \sum_{ \{ J_{l, m} \}}
 \prod_{\langle l, m \rangle} P^{({\rm D} \pm J )} (J_{l, m})  \nonumber \\
 &\times& \frac{\sum_{\{ S_l \} } 
 P_{\rm FK} (S_i, S_j, J_{i, j})
 \, e^{\beta_P \sum_{\langle l, m \rangle} J_{l, m} S_l S_m}}
 {\sum_{\{ S_l \} } 
 e^{\beta_P \sum_{\langle l, m \rangle} J_{l, m} S_l S_m}} \nonumber \\
 &=&
 \frac{1}
 {2^N (e^{\beta^{(2)}_P J^2 + \beta_P J} + 1 + e^{\beta^{(2)}_P J^2 - \beta_P J})^{N_B}}  \nonumber \\
 &\times& \sum_{ \{ J_{l, m} \}} \sum_{\{ S_l \} } 
 P_{\rm FK} (S_i, S_j, J_{ij}) \times \nonumber \\
 & & e^{\beta^{(2)}_P \sum_{\langle lm \rangle} J^2_{lm} + 
 \beta_P \sum_{\langle l, m \rangle} J_{l, m} S_l S_m} \nonumber \\
 &=& 
 (1 - r)  \tanh ( \beta_P J )
 \, . \label{eq:bbondijDpmJ}
\end{eqnarray}
 By using
 Eqs.~(\ref{eq:GaugeT}), (\ref{eq:PbondSiSjJij}), (\ref{eq:kbondiTR}), 
 and (\ref{eq:PJijdpmJ}),
 when $\beta = \beta_P$, 
 the thermodynamic quantity of the node degree for FK bonds at node $i$,
 $[\langle k_{\rm FK} (i) \rangle_T ]^{({\rm D} \pm J)}_R$, is obtained as
\begin{equation}
 [ \langle k_{\rm FK} (i) \rangle_T ]^{({\rm D} \pm J)}_R =
 (1 - r) \tanh ( \beta_P J )
 \, k (i) \, . \label{eq:kbondiDpmJ}
\end{equation}
 By using
 Eqs.~(\ref{eq:GaugeT}), (\ref{eq:PbondSiSjJij}), (\ref{eq:k2bondiTR}), 
 and (\ref{eq:PJijdpmJ}),
 when $\beta = \beta_P$, 
 the thermodynamic quantity of the square of the node degree for FK bonds at node $i$,
 $[\langle k^2_{\rm FK} (i) \rangle_T ]^{({\rm D} \pm J)}_R$, is obtained as
\begin{eqnarray}
 & & [ \langle k^2_{\rm FK} (i) \rangle_T ]^{({\rm D} \pm J)}_R = \nonumber \\
 & & (1 - r)^2 \tanh^2 ( \beta_P J )
 \, k (i) [ k (i) - 1]   \nonumber \\ 
 &+& (1 - r) \tanh ( \beta_P J ) \, k (i) \, . \label{eq:k2bondiDpmJ}
\end{eqnarray}

We set
\begin{equation}
 \rho^{({\rm D} \pm J)}  ( \beta_P ) = (1 - r) \tanh ( \beta_P J ) \, .
 \label{eq:rhodpmj}
\end{equation}
 Equations~(\ref{eq:bbondijDpmJ}), (\ref{eq:kbondiDpmJ}),
 (\ref{eq:k2bondiDpmJ}), and (\ref{eq:rhodpmj}) are formulated as
 Eqs.~(\ref{eq:bondij}), (\ref{eq:k1bondi}), and (\ref{eq:k2bondi}).
 Therefore, the magnitude of the FK bond does not depend on $k (i)$.
 By using Eqs.~(\ref{eq:kbondTR}), (\ref{eq:k2bondTR}),
 (\ref{eq:k2k1bond}), (\ref{eq:kbondiDpmJ}), and (\ref{eq:k2bondiDpmJ}),
 we obtain
\begin{equation}
 \frac{2 (1 - r)(1 - e^{- 2 \beta_P J})}
 {1 + e^{- 2 \beta_P J} + (1 - r) (1 - e^{- 2 \beta_P J})} \ge \frac{2 \langle k \rangle_N}{\langle k^2 \rangle_N}
 \, . \label{eq:bondPcDpmJ}
\end{equation}
 Equation~(\ref{eq:bondPcDpmJ}) is the inequality when 
 the FK cluster is percolated.
 Equation~(\ref{eq:bondPcDpmJ}) is the equality when 
 the FK cluster is at the percolation transition point. 

From Eqs.~(\ref{eq:rhorange}) and (\ref{eq:rhodpmj}),
 there is the percolation transition point for 
 $r \ne 1$ and $0 < \beta_P \le \infty$.
 By using Eq.~(\ref{eq:bondPcDpmJ}), we obtain
\begin{equation}
 \frac{( 2 - a) (1 - r ) - a}{(2 - a) (1 - r) + a } \ge 
 e^{- 2 \beta_P J} \ge 0
 \, . \label{eq:bondPcDpmJ2}
\end{equation}
 By using the left-hand side of Eq.~(\ref{eq:bondPcDpmJ2})
 and the right-hand side of Eq.~(\ref{eq:bondPcDpmJ2}), we obtain 
\begin{equation}
 1 - r \ge  \frac{a}{2 - a} \, .  \label{eq:bondPcDpmJ3}
\end{equation} 
 When Eq.~(\ref{eq:bondPcDpmJ3}) is satisfied,
 there is the percolation transition point.
 
By  using Eqs.~(\ref{eq:betadpmJ}) and (\ref{eq:bondPcDpmJ}), 
 the probability $p^{({\rm D} \pm J)}$ that 
 the interaction is ferromagnetic is obtained as
\begin{equation}
 p^{({\rm D} \pm J)} = \frac{(2 - a) (1 - r ) + a}{2 (2 - a) (1 - r )} 
  \label{eq:pbondDpmJ} 
\end{equation} 
 at the percolation transition point.
 By using Eqs.~(\ref{eq:betadpmJ}) and (\ref{eq:pbondDpmJ}), 
 the percolation transition temperature $T^{({\rm D} \pm J)}_P$ is 
 obtained as 
\begin{equation}
 T^{({\rm D} \pm J )}_P
 = \frac{J}{k_B}
 \frac{2}{\ln \frac{ ( 2 - a) (1 - r ) + a }
{ ( 2 - a) (1 - r ) - a}} \, .
 \label{eq:TPbondDpmJ} 
\end{equation}

\section{The Gaussian model} \label{sec:6}

The distribution $P^{({\rm Gaussian})} (J_{ij})$
 for the Gaussian model is given in Eq.~(\ref{eq:PGaussJij}).
 We set \cite{N1}
\begin{equation}
 \beta_P = \frac{J_0}{J^2} \, . \label{eq:betagauss}
\end{equation}
 When the value of $\beta_P$
 is consistent with the value of 
 the inverse temperature $\beta$, 
 the line on the phase diagram obtained using Eq.~(\ref{eq:betagauss})
 is called the Nishimori line.
 By using the gauge transformation,  
 the distribution $P^{(\pm J)} (J_{i, j})$ part
 becomes
\begin{eqnarray}
 & & \prod_{\langle i, j \rangle} P^{(\rm Gaussian)} (J_{i, j}) \nonumber \\
 &=& \frac{1}{(2 \pi J^2)^{\frac{N_B}{2}}} \,
 e^{- \frac{N_B J^2_0}{2 J^2} - \frac{1}{2 J^2} 
 \sum_{\langle i, j \rangle} J^2_{i, j} + \frac{J_0}{J^2}  \sum_{\langle i, j \rangle} J_{i, j}} 
 \nonumber \\
 &\to& 
 \frac{1}{(2 \pi J^2)^{\frac{N_B}{2}}} \,
 e^{- \frac{N_B J^2_0}{2 J^2} - \frac{1}{2 J^2} 
 \sum_{\langle i, j \rangle} J^2_{i, j} + \frac{J_0}{J^2}
 \sum_{\langle i, j \rangle} J_{i, j} \sigma_i \sigma_j} 
 \, .
 \label{eq:PGaussJij2}
\end{eqnarray}

By using
 Eqs.~(\ref{eq:GaugeT}), (\ref{eq:PbondSiSjJij}),
 (\ref{eq:bbondijTR}), (\ref{eq:betagauss}), and (\ref{eq:PGaussJij2}),
 when $\beta = \beta_P$, 
 the thermodynamic quantity of the FK bond put between
 the spins on nodes $i$ and $j$,
 $[\langle b_{\rm FK} (i, j) \rangle_T ]^{({\rm Gaussian})}_R$,
 is obtained as
\begin{eqnarray}
 & & [ \langle b_{\rm FK} (i, j) \rangle_T ]^{({\rm Gaussian})}_R \nonumber \\
 &=& \int^{\infty}_{- \infty} \cdots \int^{\infty}_{- \infty} 
 \prod_{\langle l, m \rangle} d J_{l, m} \prod_{\langle l, m \rangle} P^{({\rm Gaussian})} (J_{l, m})
  \nonumber \\
 &\times&  \frac{\sum_{\{ S_l \} } 
 P_{\rm FK} (S_i, S_j, J_{i, j})
 \, e^{\beta_P \sum_{\langle l, m \rangle} J_{l, m} S_l S_m}}
 {\sum_{\{ S_l \} } 
 e^{\beta_P \sum_{\langle l, m \rangle} J_{l, m} S_l S_m}} 
  \nonumber \\
 &=&
  \frac{1}{2^N (2 \pi J^2 )^{N_B / 2}}
  e^{- N_B \frac{J^2_0}{2 J^2}}  \nonumber \\
 &\times& \int^{\infty}_{- \infty} \cdots \int^{\infty}_{- \infty}
  \prod_{\langle l, m \rangle} d J_{l, m} 
 \sum_{\{ S_l \} } P_{\rm FK} (S_i, S_j, J_{i, j})  \nonumber \\
 &\times& e^{- \sum_{\langle l, m \rangle} \frac{J^2_{l, m}}{2 J^2} +
 \beta_P \sum_{\langle l, m \rangle} J_{l, m} S_l S_m} \nonumber \\
 &=&
 {\rm erf} (\beta_P J / \sqrt{2}) \, , \label{eq:bbondijGauss}
\end{eqnarray}
 where ${\rm erf} (x)$ is the error function of $x$.
 By using
 Eqs.~(\ref{eq:GaugeT}),
 (\ref{eq:PbondSiSjJij}), (\ref{eq:kbondiTR}), (\ref{eq:betagauss}), 
 and (\ref{eq:PGaussJij2}),
 when $\beta = \beta_P$, 
 the thermodynamic quantity of the node degree for FK bonds at node $i$,
 $[\langle k_{\rm FK} (i) \rangle_T ]^{({\rm Gaussian})}_R$, is obtained as
\begin{equation}
 [ \langle k_{\rm FK} (i) \rangle_T ]^{({\rm Gaussian})}_R =
 {\rm erf} (\beta_P J / \sqrt{2}) \, k (i) \, . \label{eq:kbondiGauss}
\end{equation}
 By using
 Eqs.~(\ref{eq:GaugeT}), (\ref{eq:PbondSiSjJij}),
 (\ref{eq:k2bondiTR}), (\ref{eq:betagauss}), and (\ref{eq:PGaussJij2}), 
 when $\beta = \beta_P$, 
 the thermodynamic quantity of the square of the node degree for FK bonds at node $i$,
 $[\langle k^2_{\rm FK} (i) \rangle_T ]^{({\rm Gaussian})}_R$, is obtained as
\begin{eqnarray}
 & & [ \langle k^2_{\rm FK} (i) \rangle_T ]^{({\rm Gaussian})}_R  \nonumber \\
 &=& [ {\rm erf} (\beta_P J / \sqrt{2}) ]^2 
  k (i) [ k(i) - 1]  \nonumber \\ 
 &+& {\rm erf} (\beta_P J / \sqrt{2}) \, k (i) \, . \label{eq:k2bondiGauss}
\end{eqnarray}

We set
\begin{equation}
 \rho^{({\rm Gaussian})} ( \beta_P ) = {\rm erf} (\beta_P J / \sqrt{2})
 \, . \label{eq:rhogauss}
\end{equation}
 Equations~(\ref{eq:bbondijGauss}), (\ref{eq:kbondiGauss}),
 (\ref{eq:k2bondiGauss}), and (\ref{eq:rhogauss}) are formulated as
 Eqs.~(\ref{eq:bondij}), (\ref{eq:k1bondi}), and (\ref{eq:k2bondi}).
 Therefore, the magnitude of the FK bond does not depend on $k (i)$.
 By using Eqs.~(\ref{eq:kbondTR}), (\ref{eq:k2bondTR}), (\ref{eq:k2k1bond}),
 (\ref{eq:kbondiGauss}), and (\ref{eq:k2bondiGauss}), we obtain
\begin{equation}
 \frac{2 \, {\rm erf} (\beta_P J / \sqrt{2})}
 { {\rm erf} (\beta_P J / \sqrt{2}) + 1} \ge
 \frac{2 \langle k \rangle_N}{\langle k^2 \rangle_N} \, . \label{eq:bondPcGauss}
\end{equation}
 Equation~(\ref{eq:bondPcGauss}) is the inequality when 
 the FK cluster is percolated.
 Equation~(\ref{eq:bondPcGauss}) is the equality when 
 the FK cluster is at the percolation transition point. 

From Eqs.~(\ref{eq:rhorange}) and (\ref{eq:rhogauss}),
 there is the percolation transition point for $0 < \beta_P \le \infty$.
 From Eq.~(\ref{eq:bondPcGauss}),
 there is the percolation transition point for $0 < a \le 1$.
 We approximate the error function 
 ${\rm erf} (x)$ by
\begin{equation}
 {\rm erf} (x) \approx
 \sqrt{1 - \exp (- 4 x^2 / \pi)} \, .
 \label{eq:erfx}
\end{equation}
 By using Eqs.~(\ref{eq:betagauss}),
 (\ref{eq:bondPcGauss}), and (\ref{eq:erfx}), 
 $J_0 / J$ is obtained as 
\begin{equation}
 \frac{J_0}{J} = 
 \sqrt{\frac{\pi}{2} \ln
 \frac{(2 - a)^2}{4 (1 - a)} } 
 \label{eq:J0JbondGauss}
\end{equation}
 at the percolation transition point.
 By using Eqs.~(\ref{eq:betagauss}) and (\ref{eq:J0JbondGauss}), 
 the percolation transition temperature
 $T^{({\rm Gaussian})}_P$ is obtained as
\begin{equation}
 T^{({\rm Gaussian} )}_P = \frac{J}{k_B}
 \frac{1}{\sqrt{\frac{\pi}{2} \ln
  \frac{(2 - a)^2}{4 (1 - a)}  }} \, . 
 \label{eq:TPbondGauss} 
\end{equation}

\begin{figure}[t]
\begin{center}
\includegraphics[width=0.95\linewidth]{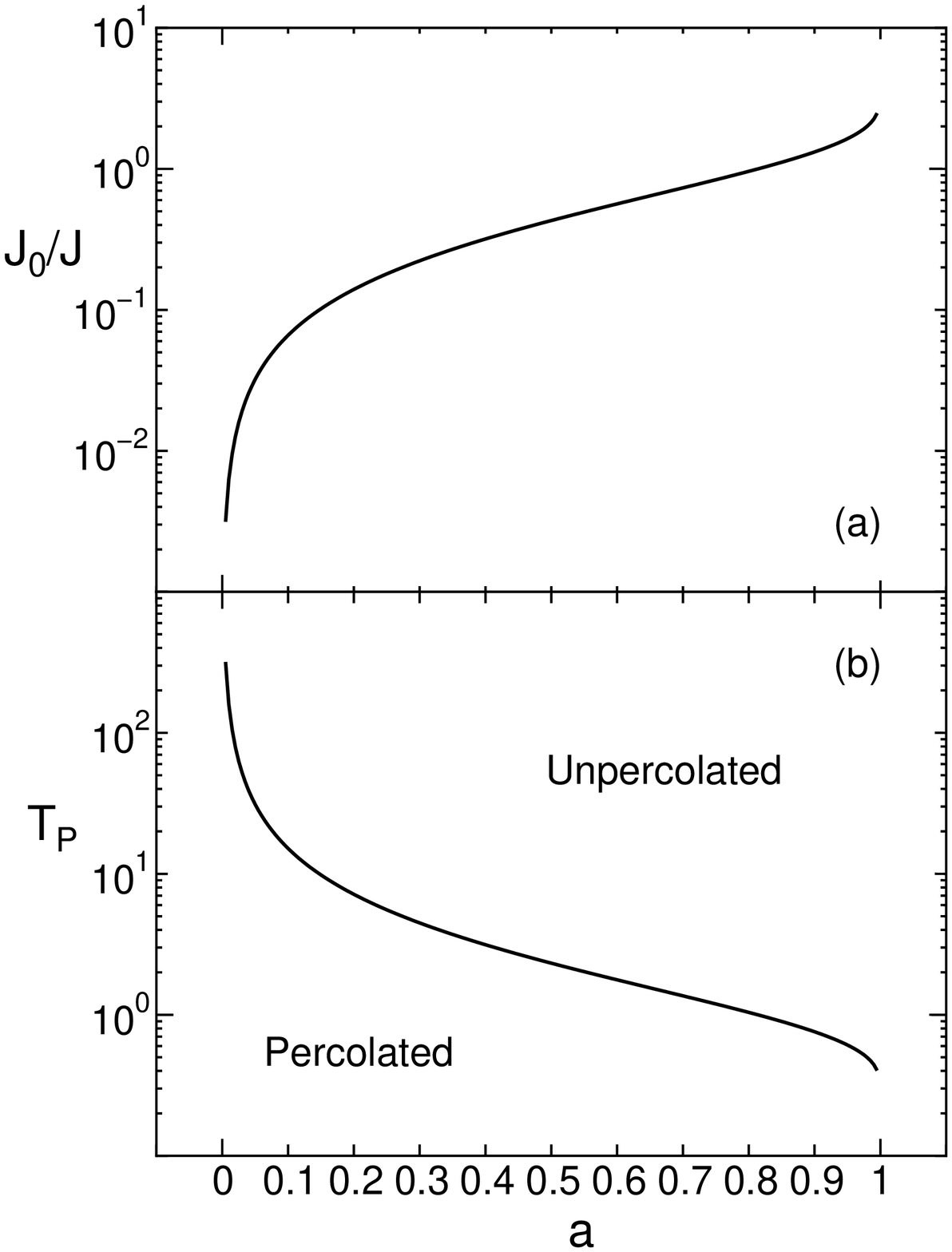}
\end{center}
\caption{
 Percolation threshold of the FK cluster for the Gaussian model.
 (a) The relation between the aspect $a$ and $J_0 / J$ is
 shown.
 (b) The relation between the aspect $a$ and the percolation
 transition temperature $T^{({\rm Gaussian})}_P$ is shown. $J / k_B$ is set to $1$.
 \label{fig:Gauss}
}
\end{figure}
 Figure~\ref{fig:Gauss} shows
 the percolation threshold of the FK cluster for the Gaussian model.
 Figure~\ref{fig:Gauss}(a) shows
 the relation between the aspect $a$ and $J_0 / J$.
 Equation~(\ref{eq:J0JbondGauss}) is used for showing Fig.~\ref{fig:Gauss}(a).
 Figure~\ref{fig:Gauss}(b) shows
 the relation between the aspect $a$ and the percolation
 transition temperature $T^{({\rm Gaussian})}_P$.
 Equation~(\ref{eq:TPbondGauss}) is used for showing Fig.~\ref{fig:Gauss}(b).
 $J / k_B$ is set to $1$.

The complete graph is considered as $a \sim 0$.
 We set
 $\langle k \rangle_N = N - 1$, $\langle k^2 \rangle_N = (N - 1)^2$,
 $a = 2 / (N - 1)$, $J_0 \to J_0 / N$, and $J \to J / \sqrt{N}$.
 From the settings, the model on the network becomes the SK model \cite{SK}.
 By using Eq.~(\ref{eq:J0JbondGauss}),
 $J_0 / J$ is obtained as
\begin{equation}
 \frac{J_0}{J} = \sqrt{\frac{\pi N}{2} 
 \ln \biggl(1 + \frac{1}{N^2 - 4 N + 3} \biggr)}
 \to \sqrt{ \frac{\pi}{2 N} }  \label{eq:J0JbondGauss2}
\end{equation}
 for a sufficiently large number of nodes
 at the percolation transition point.
 By using Eq.~(\ref{eq:TPbondGauss}),
 the percolation transition temperature $T^{({\rm SK})}_P$
 is obtained as
\begin{eqnarray}
 T^{({\rm SK})}_P &=& \frac{J}{k_B}
 \frac{1}{\sqrt{\frac{\pi N}{2} \ln 
 \biggl( 1 + \frac{1}{N^2 - 4 N + 3} \biggr) }} \nonumber \\
 &\to& \frac{J}{k_B} \sqrt{\frac{2 N}{ \pi}} 
 \label{eq:TPbondGauss2} 
\end{eqnarray}
 for a sufficiently large number of nodes.

We consider the case for $a = 1$.
 By using Eq.~(\ref{eq:J0JbondGauss}), we obtain $J_0 / J = \infty$.
 By using Eq.~(\ref{eq:TPbondGauss}), we obtain $T_P = 0$.
 From Eq.~(\ref{eq:k2k1}),
 the network is at the percolation transition point.
 From $J_0 / J = \infty$, the exchange interaction is only the ferromagnetic interaction.
 From $T_P = 0$, all the spins are parallel.
 From $J_0 / J = \infty$ and $T_P = 0$, we obtain $P_{\rm FK} = 1$
 for all nearest-neighbor pairs.
 Therefore, the FK cluster and the network are 
 at the percolation transition point. 
 We were able to confirm 
 that our result is exact at this extremal point.

In the result for the Gaussian model,
 an approximate formula for the error function,
 Eq.~(\ref{eq:erfx}), is used.
 In the result for the Gaussian model,
 it is necessary for the more precise estimation of 
 the percolation threshold that the error function in 
 Eq.~(\ref{eq:bondPcGauss}) is numerically estimated.

\section{Summary}  \label{sec:7}

In this article,
 the $\pm J$ Ising model, the diluted $\pm J$ Ising model,
 and the Gaussian Ising model on random graphs with artibary 
 degree distributions were investigated.

The values of $[\langle b_{\rm FK} (i, j) \rangle_T]_R$,
 $[\langle k_{\rm FK} (i) \rangle_T]_R$, $[\langle k^2_{\rm FK} (i) \rangle_T]_R$,
 $[\langle k_{\rm FK} \rangle_T]_R$, and $[\langle k^2_{\rm FK} \rangle_T]_R$
 on the Nishimori line were shown.
 They are quantities for the FK bonds,
 and are exact even on a finite number of nodes.

It is known that 
 the internal energy, the upper bound of the specific heat, 
 and so forth are exactly calculated
 on the Nishimori line without the dependence 
 of the network (lattice) \cite{N1, MH, N2, HM, H}.
 In this article, 
 it was realized that, as a property on the Nishimori line, 
 the magnitude of the FK bond does not depend on the degree $k (i)$.

The percolation thresholds of the FK cluster were shown.
 We used a conjectured criterion
 (Eq.~(\ref{eq:k2k1bond})) to obtain the thresholds.
 We were able to confirm that 
 our results are exact at several extremal points.
 Therefore, our entire set of results may be exact.

\end{document}